\documentclass[pra,prl,twocolumn,showpacs,floatfix,10pt]{revtex4-1}
\usepackage{graphicx}
\usepackage{dcolumn}
\usepackage{bm}

\usepackage{amsmath}
\usepackage{latexsym,amssymb,graphicx}
\usepackage[mathscr]{eucal}
\usepackage[bookmarks=true,
   colorlinks=true,
   linkcolor=blue,
   urlcolor=blue,
   citecolor=blue,
   bookmarks=true,
   hyperindex=true
]{hyperref}

\begin{document}


\title{Constraining the generalized uncertainty principle with the atomic weak-equivalence-principle test}

\author{Dongfeng Gao\textsuperscript{1,2,}}
\altaffiliation{Email: dfgao@wipm.ac.cn}
\author{Jin Wang\textsuperscript{1,2}}
\author{Mingsheng Zhan\textsuperscript{1,2,} }
\altaffiliation{Email: mszhan@wipm.ac.cn}
\vskip 0.5cm
\affiliation{1 State Key Laboratory of Magnetic Resonance and Atomic and Molecular Physics, Wuhan Institute of Physics and Mathematics, Chinese Academy of Sciences, Wuhan 430071, China\\
2 Center for Cold Atom Physics, Chinese Academy of Sciences, Wuhan 430071, China }

\date{\today}

\begin{abstract}
Various models of quantum gravity imply the Planck-scale modifications of Heisenberg's uncertainty principle into a so-called generalized uncertainty principle (GUP). The GUP effects on high-energy physics, cosmology, and astrophysics have been extensively studied. Here, we focus on the weak-equivalence-principle (WEP) violation induced by the GUP. Results from the WEP test with the $^{85}$Rb-$^{87}$Rb dual-species atom interferometer are used to set upper bounds on parameters in two GUP proposals. A $10^{45}$-level bound on the Kempf-Mangano-Mann proposal, and a $10^{27}$-level bound on Maggiore's proposal, which are consistent with bounds from other experiments, are obtained. All these bounds have huge room for improvement in the future.

\end{abstract}

\maketitle

\section{Introduction}

Einstein's theory of general relativity is a well-tested classical theory of gravitation, which is based on the equivalence principle \cite{will2014}. For many years, people have been working on building a quantum theory of gravity (also called quantum gravity). Despite lots of attempts, establishing quantum gravity is still unsuccessful, and has become one of the main challenges in modern physics. To gain insights into developing such quantum theories, it is useful to study experimentally accessible quantum gravity effects. Normally, such effects are tiny, because they are expected to be inversely proportional to the Planck energy scale $E_P= \sqrt{\hbar c^5 /G}=1.2 \times 10^{19}$ GeV. Even so, current experiments, mainly from the high-energy physics, astrophysics, and cosmology, can be used to constrain parameters in many models of quantum gravity. A lot of work has been done on this aspect \cite{camelia2013}.

Among the numerous quantum gravity effects, one effect is of particular importance, the so-called generalized uncertainty principle (GUP). It is well known that Heisenberg's uncertainty principle (HUP) lies at the heart of quantum mechanics. According to the HUP, uncertainties in the measurement of the length and the momentum satisfy the relation $\Delta x \Delta p \geq |\langle[x,p]\rangle|/2=\hbar/2$. In other words, the uncertainty $\Delta x$ is bounded by $\Delta x \geq \hbar /(2\Delta p)$. Therefore, it is clear that, upon the loss of information on the momentum, the length can be arbitrarily precisely measured. On the other hand, various models on quantum gravity, such as string theory, predict the existence of a minimum measurable length \cite{maggiore1993,gross1987,amati1989,konishi1990}. Thus, to be consistent with quantum gravity, the HUP should be modified into the GUP.

The GUP effects on a wide range of physical systems have been extensively investigated. For example, in high-energy physics, the GUP effects on the quark-gluon plasma were studied in Ref. \cite{elmashad2014}. The impact of the GUP on thermodynamical parameters and the stability of the Schwarzschild black hole was investigated in Ref. \cite{sabri2012}. The GUP effects on quantities of the inflationary dynamics and the thermodynamics of the early universe were calculated in Ref. \cite{tawfik2013}. The GUP corrections to the Lamb shift, the Landau levels, and the tunneling current in a scanning tunneling microscope were discussed in Refs. \cite{das2008, das2011}. The applications of the GUP in several macroscopic systems, such as the gravitational-wave bar detectors \cite{gwbar2013} and the macroscopic harmonic oscillators \cite{bawaj2015}, were reported. Furthermore, the violation of the equivalence principle induced by the GUP was discussed in Refs. \cite{tkachuk2013,ghosh2013}. More references can be found in Ref. \cite{tawfik2014}.

In recent years, rapid technological progress in atom interferometry has been made, which provides us a new tool with which to study the GUP. Because of their high sensitivity, atom interferometers have already been used in various precision measurements. Many impressive experimental results have been achieved. For example, the value of the fine structure constant, $\alpha$, was determined to be $\alpha^{-1}$=137.035999037(91) in the $^{87}$Rb-atom recoil experiment \cite{clade2011}, which is the second best value compared to the one deduced from the electron anomaly measurement \cite{codata2014}. Using a double atom-interferometer-gravity-gradiometer, the Newtonian gravitational constant was measured to be $G=6.67191(99)\times 10^{-11} {\rm m}^3 \, {\rm kg}^{-1}{\rm s}^{-2}$ \cite{tino2014}, which was one of the 14 measured values adopted by the 2014 CODATA adjustment \cite{codata2014}. A $10^{-8}$-level test of the weak equivalence principle (WEP) using the dual-species atom interferometer was reported in a recent work \cite{zhan2015}, which is the best result in the quantum test of the WEP with microscopic objects. More details on atom interferometers can be found in the review paper \cite{cronin2009}.

Having these achievements with atom interferometry in hand, people began to investigate the possibility of searching quantum gravity effects in atomic physics. One such study was given in Ref. \cite{tino2009}, where the authors used results from the $^{133}$Cs-atom recoil experiment \cite{wicht2002} to constrain parameters in the quantum-gravity-modified energy-momentum dispersion relation. Motivated by their work, we reanalyzed results of the $^{87}$Rb-atom recoil experiment \cite{clade2011}, and obtained important upper bounds on parameters in three GUP proposals \cite{gao2016}.

In this paper, we study the possibility of using the dual-species atom interferometer to search for the WEP violation induced by the GUP, and thus we constrain parameters in two popular GUP proposals. The paper is organized as follows. In Sec. II, the WEP violation induced by the GUP is discussed, and two popular GUP proposals are introduced. Then a brief description of the WEP test with dual-species atom interferometers is given in Sec. III. Through a detailed calculation of the GUP effects on the WEP test with dual-species atom interferometers, bounds on the GUP parameters are obtained in Sec. IV. Finally, we conclude the paper in Sec. V.

\section{The WEP violation and the GUP proposals}

\subsection{The WEP violation by the GUP}

The WEP states that the trajectory of a freely falling test body does not depend on its internal structure and composition. In other words, if one drops two different bodies in a gravitational field, then they will fall with the same acceleration. Thus, the WEP is also called the universality of free fall \cite{will2014}.

To illustrate the WEP violation induced by the GUP, let us assume a generic form of the GUP,
\begin{equation}
[x_i, p_j]=i \hbar f_{ij}(\vec{p}),
\label{gup}
\end{equation}
where $f_{ij}(\vec{p})$ is a symmetric function of $\vec{p}$. If $f_{ij}(\vec{p})$ is taken to be $\delta_{ij}$, then the GUP goes back to the HUP.

The way to show the WEP violation by the GUP is to consider the Heisenberg equation of motion. For our purpose, it is enough to work with the one-dimensional case,
\begin{eqnarray}
\nonumber  \dot{x} &=& \frac{1}{i \hbar} [x, H],\\
\dot{p}_x &=& \frac{1}{i \hbar} [p_x, H],
\label{eom}
\end{eqnarray}
where the Hamiltonian is $H=p_x^2/(2 m)+ m g x$.
Applying the GUP (\ref{gup}) to Eq. (\ref{eom}), we have
\begin{eqnarray}
\nonumber  \dot{x} &=& \frac{p_x}{m}f(p_x),\\
\dot{p}_x &=& -m g f(p_x).
\label{eom1}
\end{eqnarray}
It is obvious that the WEP is satisfied for the HUP. Since $f(p_x)$ is a non-trivial function for the GUP, then the WEP is violated. In other words, the trajectory, determined in Eq. (\ref{eom1}), depends on the test body's mass through the non-trivial dependence of $p_x$.

The physical explanation for the WEP violation is as follows. By its statement, the WEP is local in nature. On the other hand, the GUP is always associated with a minimum length scale, which implies that the GUP is non-local. So the WEP should be violated by the GUP, although the violation is tiny \cite{ghosh2013}.

\subsection{Two proposals on the GUP}

As discussed before, various models of quantum gravity, such as string theory and loop quantum gravity, suggest the existence of a minimum measurable length. This fact in turn indicates that the HUP should be modified into the GUP. Unfortunately, at the moment no model has the ability to predict exactly what the GUP should be. In other words, we do not know the exact form of $f_{ij}(\vec{p})$ in Eq. (\ref{gup}). An alternative way is making proposals for the GUP. Depending on what indications from models on quantum gravity are to be incorporated, various GUP proposals have been put forward \cite{tawfik2014}. Here, we discuss two popular ones.

First, let us introduce the so-called Kempf-Mangano-Mann (KMM) proposal, which is put forward in Ref. \cite{kmm1995}. The motivation behind this proposal is the observation that a variety of models on quantum gravity predicted a leading quadratic-in-the-momenta type correction to the HUP. Then, the following form is proposed
\begin{equation}
[x_i, p_j]=i \hbar \left(\delta_{ij} + \frac{\beta_0}{(M_P c)^2} \delta_{ij} p^2 + \frac{2\beta_0}{(M_P c)^2} p_i p_j\right),
\label{kmm}
\end{equation}
where $p^2:=\vec{p}^2=\sum_{j=1}^3 p^j p_j$, $\beta_0$ is a dimensionless parameter, and $M_P = \sqrt{\hbar c /G}$ is the Planck mass. All other commutation relations vanish. Accordingly, one has the following uncertainty relation:
\begin{equation}
\Delta x_i \Delta p_i \geq \frac{\hbar}{2} \left(1 + \frac{\beta_0}{(M_P c)^2}\left((\Delta p)^2+\langle p \rangle^2 + 2 \Delta p_i^2 +2\langle p_i \rangle^2 \right) \right).
\end{equation}
This inequality relation implies a minimum measurable length $\Delta x_{min}=\sqrt{3\beta_0}L_P$, where $L_P=\sqrt{\hbar G /c^3}$ is the Planck length. Normally, the Planck length is believed to be the minimal measurable length. Thus, $\beta_0$ is assumed to be of the order of unity. However, if one does not take the above assumption \textit{a priori}, current experiments can be used to set bounds on $\beta_0$. For example, the standard model of high-energy physics is well tested at an energy scale of 100 GeV, which implies that $\beta_0 \leq 10^{34}$. Better bounds are obtained in Refs. \cite{das2008, bawaj2015}, where the best one is set by macroscopic harmonic oscillators.

Next, let us consider Maggiore's proposal \cite{maggiore1993,maggiore1995}. It is motivated by the study of the relationship between the GUP and the quantum deformation of the Poincar${\rm\acute{e}}$ algebra. The form of the proposal is
\begin{equation}
[x_i, p_j]=i \hbar \, \delta_{ij}\sqrt{1+\frac{\gamma_0}{(M_P c)^2}(p^2+m^2 c^2)},
\label{maggiore1}
\end{equation}
where $\gamma_0$ is a dimensionless parameter, and is normally assumed to be of the order of unity. If this assumption is not taken \textit{a priori}, many experiments can be used to set upper bounds on it. The associated minimum measurable length is found to be $\Delta x_{min}\simeq\sqrt{\gamma_0/2} L_P$. Again, a direct bound from high-energy physics is $\gamma_0 \leq 10^{34}$. Since $\gamma_0/(M_P c)^2$ is very small, one can make a first-order Taylor expansion on Eq. (\ref{maggiore1}):
\begin{equation}
[x_i, p_j]=i \hbar \, \delta_{ij}\left(1+\frac{\gamma_0}{2(M_P c)^2}(p^2+m^2 c^2)\right).
\end{equation}
Compared to the KMM proposal, the above formula is very similar to it. But the difference will be important for later experimental analysis.

\section{The WEP test with atom interferometers}

In this section, we first review the theory of Raman atom interferometers. Then, we briefly describe how to use atom interferometers to test the WEP.

\subsection{Theory of Raman atom interferometers}

\begin{figure}[htbp]
\centering
\includegraphics[width=10.0cm]{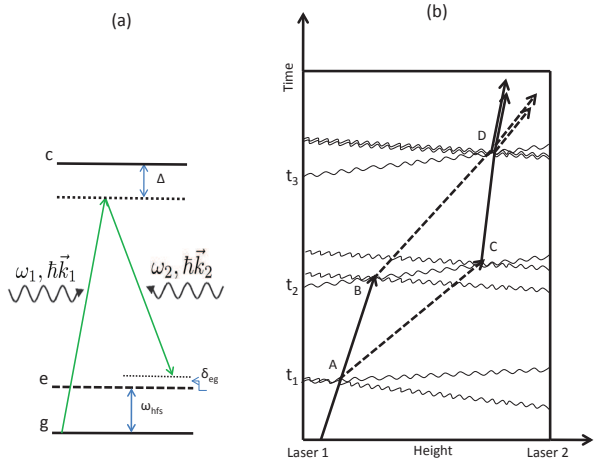}
\caption{Schematic diagrams for a $\pi/2$-$\pi$-$\pi/2$ Raman atom interferometer. (a) Diagram for Raman transitions, which happen between two hyperfine ground states $|g\rangle$ and $|e\rangle$. Laser beams are detuned by $\Delta$ from the optical resonance $|c\rangle$. The atomic population is resonantly transferred between $|g\rangle$ and $|e\rangle$ when the frequency difference $\omega_1-\omega_2$ is close to $\omega_{hfs}$. (b) Diagram for paths of an atom in the interferometer, where Raman $\pi/2$ pulses are applied at points A and D, and Raman $\pi$ pulses are applied at points B and C. }
\label{fig1}
\end{figure}

The theory of Raman atom interferometers can be found in many papers, such as Ref. \cite{chu1992}. A typical Raman atom interferometer can be described with the schematic diagram in Fig. 1, where the Raman transitions and paths of an atom in the interferometer are shown in Figs. 1(a) and 1(b), respectively. A slow atomic wave packet, prepared in the $|g\rangle$ state with momentum $\vec{p}$ (denoted by $|g, \vec{p}\rangle$), is loaded into the interferometer. At point A, the first Raman $\pi$/2 pulse is applied to coherently split the atomic wave packet into a superposition of states $|g, \vec{p}\rangle$ and $|e,\vec{p}+\hbar\vec{k}_1-\hbar\vec{k}_2\rangle$. After a drift time $T$, Raman $\pi$ pulses are applied at points B and C, which transit the state $|g, \vec{p}\rangle$ to $|e,\vec{p}+\hbar\vec{k}_1-\hbar\vec{k}_2\rangle$ and the state $|e,\vec{p}+\hbar\vec{k}_1-\hbar\vec{k}_2\rangle$ to $|g, \vec{p}\rangle$, respectively. After another drift time $T$, the two wave packets overlap at point D. Then, the final Raman $\pi$/2 pulses are applied to make the two wave packets interfere. The interference can be detected by measuring the number of atoms in either $|g\rangle$ or $|e\rangle$ states.

The Raman transition in Fig. 1(a) involves two counter-propagating photons and three internal energy states: two hyperfine ground states $|g\rangle$ and $|e\rangle$, and a virtual excited state $|c\rangle$. Each photon is detuned by a frequency $\Delta$ from the allowed optical transition to state $|c\rangle$. The Hamiltonian of this three-level system is
\begin{equation}
H_{a}=\frac{p^2}{2m}+\hbar \omega_g |g\rangle \langle g| +\hbar \omega_e |e\rangle \langle e|+\hbar \omega_c |c\rangle \langle c|+V_{in},
\label{hamila}
\end{equation}
where $m$ is the atomic mass, and $\hbar \omega_{\alpha}$ ($\alpha=g,e,c$) is the internal energy of state $|\alpha \rangle$. $\hbar \omega_{hfs}$ is defined to be the energy difference $\hbar (\omega_{e}-\omega_{g})$. $V_{in}$ is the atom-photon interaction, which is taken to be the electric dipole coupling:
\begin{eqnarray}
\nonumber V_{in} &=& - \vec{\mu} \cdot \vec{E}(\vec{x}),\\
\nonumber \vec{E}(\vec{x}) & =& \vec{E}_1\, cos(\vec{k}_1\cdot \vec{x}-\omega_1 t +\phi_1) \\
 & & + \vec{E}_2\, cos(\vec{k}_2\cdot \vec{x}-\omega_2 t +\phi_2),
\end{eqnarray}
where $\vec{\mu}$ is the electric dipole moment operator, and $\vec{E}_1$  and $\vec{E}_2$ are fields for the two counter-propagating photons (with frequencies $\omega_1$ and $\omega_2$, and $\vec{k}_1 \approx -\vec{k}_2$ ) propagating along $\vec{x}$.

To describe the atom's center-of-mass motion, we write down the Schr${\rm \ddot{o}}$dinger equation for the above Hamiltonian, and look for solutions of the form
\begin{equation}
|\psi\rangle= \int d^3 p \sum_{\alpha=g,e,c}a_{\alpha,\vec{p}}(t){\rm e}^{i(\frac{p^2}{2m}+\omega_\alpha)t}|\alpha,\vec{p}\rangle,
\end{equation}
where $|\alpha,\vec{p}\rangle$ stands for an atom in internal state $|\alpha\rangle$ and momentum eigenstate $\psi_{\vec{p}}(\vec{x})$ ($\sim e^{i \vec{p}\cdot \vec{x}/\hbar}$). In fact, this three-level system can be reduced to a two-level system after adiabatic elimination of state $|c\rangle$.

With some efforts, the following solutions can be found. For a Raman $\pi$/2 transition of duration $\tau$, the solutions are
\begin{equation*}
a_{g,\vec{p}}(t+\tau) = \sqrt{\frac{1}{2}}\, [a_{g,\vec{p}}(t)-i {\rm e}^{i \phi(t)}a_{e,\vec{p}+\hbar\vec{k}_1-\hbar\vec{k}_2}(t)],
\end{equation*}
\begin{eqnarray}
a_{e,\vec{p}+\hbar\vec{k}_1-\hbar\vec{k}_2}(t+\tau) &=& \sqrt{\frac{1}{2}}\, [-i {\rm e}^{-i \phi(t)}a_{g,\vec{p}}(t)\\
 \nonumber  & &+ a_{e,\vec{p}+\hbar\vec{k}_1-\hbar\vec{k}_2}(t)],
\end{eqnarray}
where
\begin{equation}
\phi(t)=\int_{t_0}^{t}\delta_{eg} d t'.
\label{phase}
\end{equation}
Here, $\delta_{eg}$ is the effective two-photon detuning, which is
\begin{equation}
 \delta_{eg}=\omega_1 - \omega_2 -\omega_{hfs}- \frac{\hbar}{2 m}(\vec{k}_1-\vec{k}_2)^2-(\vec{k}_1-\vec{k}_2)\cdot \vec{v}_a,
\label{sel}
\end{equation}
where $\vec{v}_a$ is the atomic velocity.

For a Raman $\pi$ transition of duration $2\tau$, the solutions are
\begin{eqnarray}
 \nonumber a_{g,\vec{p}}(t+2\tau) &= &-i {\rm e}^{i \phi(t)}a_{e,\vec{p}+\hbar\vec{k}_1-\hbar\vec{k}_2}(t),\\
a_{e,\vec{p}+\hbar\vec{k}_1-\hbar\vec{k}_2}(t+2\tau) &=& -i {\rm e}^{-i \phi(t)}a_{g,\vec{p}}(t).
\end{eqnarray}

During the drift time $T$ between two Raman pulses, the free evolution is assumed as follows:
\begin{eqnarray}
 \nonumber a_{g,\vec{p}}(t+T) &= & a_{g,\vec{p}}(t),\\
a_{e,\vec{p}+\hbar\vec{k}_1-\hbar\vec{k}_2}(t+T) &=& a_{e,\vec{p}+\hbar\vec{k}_1-\hbar\vec{k}_2}(t).
\end{eqnarray}

Then, it is straightforward to write down the phase shift for the $\pi/2$-$\pi$-$\pi/2$ Raman atom interferometer in Fig.1. Suppose the atomic beam is initially prepared in the $|g\rangle$ state. After the interference, the probability amplitudes for staying in state $|g\rangle$ and transiting
to state $|e\rangle$ are
\begin{eqnarray}
 \nonumber \mathcal{A}_{g\rightarrow g} &= &\frac{1}{2} (1+ {\rm e}^{-i \Delta\phi}),\\
\mathcal{A}_{g\rightarrow e} &= &\frac{i}{2} (1- {\rm e}^{i \Delta\phi}),
\end{eqnarray}
where $\Delta\phi$ is the phase shift, which is
\begin{equation}
 \Delta\phi=\phi(t_1)-2\phi(t_2)+\phi(t_3).
\end{equation}
Here $t_1$ is the time for the first Raman $\pi$/2 transition, $t_2=t_1+\tau +T$ is the time for the Raman $\pi$ transitions, and $t_3=t_1+3\tau +2T$ is the time for the final Raman $\pi$/2 transitions.

\subsection{Experiment of the WEP test with atom interferometers}

Most experiments on the WEP test are done with macroscopic objects. On the other hand, it is also very important to test the WEP with microscopic objects. The atom interferometer is an ideal tool for this purpose because of its high precision in measuring the gravitational acceleration.

Now, consider the above $\pi/2$-$\pi$-$\pi/2$ Raman atom interferometer in a gravitational field $\vec{g}$. The Hamiltonian is then given by
\begin{equation}
H=H_a + m \vec{g} \cdot \vec{x}.
\label{hamil2}
\end{equation}
Repeating the previous derivation, one can find the phase shift to be
\begin{equation}
 \Delta\phi=\phi_A(t_1)-\phi_B(t_2)-\phi_C(t_2)+\phi_D(t_3),
 \label{phaseshift}
\end{equation}
where $\phi_{point}(t_i)$ is still given by Eq. (\ref{phase}), and the subscripts (A,B,C,D) are only used to distinguish the phase at different point. Then, a short calculation of Eq. (\ref{phaseshift}) shows that
\begin{equation}
 \Delta\phi=-(\vec{k}_1-\vec{k}_2)\cdot \vec{g} T^2,
 \label{phaseshifta}
\end{equation}
where the short pulse limit, $\tau \rightarrow 0$, has been taken.
This formula is the foundation for a precision test of the WEP with atom interferometers.

To test the WEP, one has to drop two test bodies simultaneously. Then, the technique of so-called dual-species atom interferometry is developed, where two different atomic species are used. They can be either isotopic atoms (such as the $ ^{85}{\rm Rb}$-$ ^{87}{\rm Rb}$ pair used in Refs. \cite{fray2004,bonnin2013,zhan2015}, and the $ ^{87}{\rm Sr}$-$ ^{88}{\rm Sr}$ pair used in Ref. \cite{tarallo2014}) or nonisotopic atoms (such as the $ ^{87}{\rm Rb}$ and $ ^{39}{\rm K}$ atoms used in Ref. \cite{rasel2014}). By the simultaneous realization of Raman interference for both species, the gravitational acceleration of two atomic species can be measured simultaneously. By comparing the difference in the values of the gravitational acceleration of two different atomic species, one can test the WEP. Denote the measured values of the gravitational acceleration for atomic species 1 and 2 by $g_1$ and $g_2$, respectively. The E${\rm \ddot{o}}$tv${\rm \ddot{o}}$s parameter for the WEP test is defined to be
\begin{equation}
 \eta=\frac{g_1-g_2}{(g_1+g_2)/2}
\end{equation}

The best result for $\eta$ with microscopic objects was given in Ref. \cite{zhan2015}, where a simultaneous $^{85}$Rb-$^{87}$Rb dual species atom interferometer of drift time $T=70.96$ ms was adopted. To simultaneously realize the Raman interference for $^{85}$Rb and $^{87}$Rb atoms, the experiment proposed a four-wave double-diffraction Raman transition scheme, and demonstrated its ability in suppressing common-mode phase noise. After a careful correction of various systematic errors, the final value for $\eta$ was measured to be
\begin{equation}
 \eta=(2.8\pm 3.0)\times 10^{-8}.
 \label{hmeas}
\end{equation}

In the next section, we will discuss how to use this measurement to constrain parameters in the previous GUP proposals.

\section{bounds on the GUP parameters}

We first make some remarks on how the GUP affects the above calculation of phase shift. At each Raman transition, the atom acquires a phase through interacting with laser pulses, which satisfies both the momentum conservation law and the energy conservation law. The GUP will modify the momentum kick that the atom receives from laser pulses. Although no study has been done for rubidium atoms, we can make some rough estimation, based on the work done for GUP effects on the energy spectrum of hydrogen atoms \cite{brau1999}. The GUP modification of the energy level is of the order of $\beta_0  \frac{m_e^2}{M_P^2}  \alpha^2 E_0\sim \beta_0 \times 10^{-48}$ eV, where $m_e$ is the electronic mass, and $E_0=13.6$ eV. Obviously, this GUP modification in momentum kick, $\beta_0 \times 10^{-48}$ eV/$c$, is far less than the measurement error, $h  \Delta \nu/c\sim 10^{-12}eV$/$c$. Even if this term is kept in the following calculation, compared to the dominating term $\beta_0 \frac{m^2}{M_P^2}$, its contribution can still be ignored.

Another remark is on the derivation of the phase shift in the above section. The phase shift is derived in terms of the mean momentum of atomic wave packets in the short pulse limit. So the derivation is a semiclassical approximation in nature, which was adopted for most experiments. On the other hand, according to Refs. \cite{jansen2008,hu2015}, a full quantum mechanical calculation of the phase shift with finite Raman pulse duration is needed only for an accuracy higher than $10^{-10}$. Moreover, the quantum mechanical calculation in the short pulse limit is equivalent to the approach based on the classical trajectories.

Thus, for our purpose in this paper, it is enough to take a semiclassical treatment in constraining parameters in the GUP proposals with atom interferometers. The mean trajectory and the mean momentum of atomic wavepackets satisfy the Heisenberg equation of motion. The GUP effects are incorporated through their modification on the atom's mean trajectory and mean momentum in a gravitational field. This modification in the atom's path results in a phase shift in the interferometer, which then produces an atomic-mass-dependent value of the gravitational acceleration $g$. Therefore, a non-zero value for $\eta$ is induced by the GUP.

\subsection{Bound on the KMM proposal}

Applying the KMM proposal (\ref{kmm}) to the one-dimensional Heisenberg equation of motion (\ref{eom}), we can find
\begin{eqnarray}
\nonumber  \dot{x} &=& \frac{p_x}{m}(1+\frac{3\beta_0}{(M_P c)^2}p_x^2),\\
\dot{p}_x &=& -m g (1+\frac{3\beta_0}{(M_P c)^2}p_x^2).
\label{eomkmm}
\end{eqnarray}
To first order in $\beta_0/(M_P c)^2$, the solution to the above differential equation is
\begin{equation}
x(t)= x_0 + v_0 t - \frac{1}{2}g t^2 (1+\frac{2\beta_0}{(M_P c)^2}m^2 g^2 t^2),
\end{equation}
where $x_0$ and $v_0$ stand for the initial position and velocity of the atoms, respectively.

Once we have the GUP-modified solution to the atom's path, we can recalculate the phase shift (\ref{phaseshift}) of the $\pi/2$-$\pi$-$\pi/2$ Raman atom interferometer.  A short calculation shows that
\begin{equation}
 \Delta\phi=-(k_1+k_2)g T^2 (1+\frac{2\beta_0}{(M_P c)^2}m^2 g^2 T^2).
 \label{phaseshift1}
\end{equation}
Then, the gravitational accelerations for atomic species 1 and 2 in a dual species atom interferometer are
\begin{eqnarray}
\nonumber  g_1 &=& g (1+\frac{2\beta_0}{(M_P c)^2}m_1^2 g^2 T^2),  \,\,\, {\rm and}\\
g_2 &=& g (1+\frac{2\beta_0}{(M_P c)^2}m_2^2 g^2 T^2).
\end{eqnarray}
So, the E${\rm \ddot{o}}$tv${\rm \ddot{o}}$s parameter $\eta$ is found to be
\begin{equation}
\eta = 2\beta_0  \frac{m_1^2-m_2^2}{M_P^2} \frac{g^2 T^2}{c^2}.
\label{kmm2}
\end{equation}

Therefore, with the values in Table I, the measurement (\ref{hmeas}) from the simultaneous $^{85}$Rb-$^{87}$Rb dual-species atom interferometer sets an upper bound,
\begin{equation}
\beta_0 < 2.6 \times 10^{45}.
\end{equation}
This bound on $\beta_0$ is weaker than those set by high-energy physics, measurements of the Lamb shift \cite{das2008}, and macroscopic harmonic oscillators \cite{bawaj2015}. The reason is that $\beta_0$ is always associated with two factors, $(m_1^2-m_2^2)/M_P^2$ and $v^2/c^2$, in Eq. (\ref{kmm2}). Even if one atomic species is much heavier than the other, the factor $(m_1^2-m_2^2)/M_P^2$ could not be substantially greater than $10^{-34}$. As for the factor $v^2/c^2$, it is at most of order $10^{-14}$ for cold atoms. Taken these two limitations into consideration, it is hopeless for experiments of the WEP test with atom interferometers to give a better bound on $\beta_0$ than the one from macroscopic harmonic oscillators. However, things are different for Maggiore's proposal.

\begin{table}
\caption{Quantities used in our calculation}
\begin{tabular}{|l|l|c|}
  \hline
  Quantity & Value & Source \\ \hline
  $ g $ & 9.789 $m/s^2$ (at Wuhan) &  \cite{zhan2011} \\
  $ c $ & 299792458 $m/s$ &  \cite{codata2014} \\
  $m(^{85}Rb) $ & 84.911789739(9) $ m_u$ &  \cite{mount2010} \\
 $m(^{87}Rb) $ & 86.909180535(10) $ m_u$ &  \cite{mount2010} \\
  \hline
\end{tabular}
\end{table}

\subsection{Bound on Maggiore's proposal}

As above, applying Maggiore's proposal (\ref{maggiore1}) to Eq. (\ref{eom}), we get the following Heisenberg equation of motion:
\begin{eqnarray}
\nonumber  \dot{x} &=& \frac{p_x}{m}(1+\frac{\gamma_0}{2(M_P c)^2}(m^2 c^2+p_x^2)),\\
\dot{p}_x &=& -m g (1+\frac{\gamma_0}{2(M_P c)^2}(m^2 c^2+p_x^2)).
\label{eommaggiore}
\end{eqnarray}
Actually, since $p_x^2 \ll m^2 c^2$ for cold atoms, we can omit terms proportional to $\gamma_0 p_x^2/(M_P c)^2$. To first order in $\gamma_0/(M_P c)^2$, the solution to the atomic trajectory is
\begin{equation}
x(t)= x_0 + v_0 t - \frac{1}{2}g t^2 (1+\frac{\gamma_0}{M_P^2}m^2).
\end{equation}

With this GUP modified atomic path, the phase shift (\ref{phaseshift}) of a $\pi/2$-$\pi$-$\pi/2$ Raman atom interferometer is calculated to be
\begin{equation}
 \Delta\phi=-(k_1+k_2)g T^2 (1+\frac{\gamma_0}{M_P^2}m^2).
 \label{phaseshift1}
\end{equation}
Accordingly, the gravitational accelerations for atomic species 1 and 2 in a dual-species atom interferometer are
\begin{eqnarray}
\nonumber  g_1 &=& g (1+\frac{\gamma_0}{M_P^2}m_1^2),  \,\,\, {\rm and}\\
g_2 &=& g (1+\frac{\gamma_0}{M_P^2}m_2^2).
\end{eqnarray}
Then, the parameter $\eta$ is
\begin{equation}
\eta = \gamma_0 \cdot \frac{m_1^2-m_2^2}{M_P^2}.
\label{maggiore2}
\end{equation}

With the values in Table I, the measurement (\ref{hmeas}) gives an upper bound,
\begin{equation}
\gamma_0 < 4.0 \times 10^{27}.
\end{equation}
Clearly, this bound on $\gamma_0$ is much better than the one set by high-energy physics, and slightly worse than the one given in Ref. \cite{gao2016}. Compared to the case for the KMM proposal, the result is much better. The reason is that $\gamma_0$ is only associated with the factor $(m_1^2-m_2^2)/M_P^2$ in Eq. (\ref{maggiore2}). Without the suppression due to the factor $v^2/c^2$, the bound on $\gamma_0$ is almost 18 orders better than the one on $\beta_0$.

Here, it is important to make some clarification to the paper \cite{ghosh2013}, where the author used the framework of the geodesic equation in Einstein's theory of generality relativity to study the WEP violation induced by the GUP. First, we reexamine the author's derivation, and find that the result in Ref. \cite{ghosh2013} is very similar to our result, Eq. (\ref{maggiore2}). In fact, instead of the relativistic KMM proposal he had thought, the author was discussing Maggiore's proposal in deed. Therefore, his result was not a bound on $\beta_0$, but a bound on $\gamma_0$. Secondly, as pointed out by the author himself, the derivation in Ref. \cite{ghosh2013} did not hold for the WEP test with macroscopic bodies. However, his bound on $\beta_0$ was based on the result of the WEP test with a torsion pendulum, which is quite controversial. As comparison, our bound on $\gamma_0$ comes from the result of the WEP test with atoms, which is a more appropriate choice.

\section{Conclusion}

We have discussed the WEP violation induced by the GUP, and investigated how to use results from atomic WEP tests to constrain parameters in two popular GUP proposals. The main advantage of taking the WEP test with the $^{85}$Rb-$^{87}$Rb dual-species atom interferometer is that all the physical quantities involved can be measured with very high precision. Our calculation shows that the WEP violation induced by the KMM proposal is velocity dependent. Thus, compared to bounds set from other experiments, our bound on the KMM proposal is worse because of the suppression by the factor $v^2/c^2$. In other words, cold-atom experiments are not useful in constraining effects of the KMM proposal. On the other hand, the WEP violation induced by Maggiore's proposal is velocity independent, which is the reason why a better bound on Maggiore's proposal can be obtained. Although all these bounds are not as good as bounds from other experiments, they are consistent with each other. In addition, the WEP test with dual-species atom interferometers has huge potential in precision. With larger photon momentum transfer and longer drift time $T$, an experimental proposal of precision of one part in $10^{17}$ was suggested in Ref. \cite{kasevich2007}. This means that a more than 10 orders of magnitude improvement on bounds of the two GUP proposals can be expected in the future, which is very impressive.

\section*{Acknowledgements}

This work was supported by the National Key Research Program of China under Grant No. 2016YFA0302002, the National Science Foundation of China under Grants No. 11227803 and No. 91536221, and the Strategic Priority Research Program of the Chinese Academy of Sciences under Grant No. XDB21010100.

\end{document}